\documentstyle[prl,aps,epsf,multicol]{revtex}
\begin{document}

%
%



\title{Fractionalized Fermi liquids}

\author{T. Senthil$^{(1)}$, Subir Sachdev$^{(2)}$, and Matthias Vojta$^{(3)}$}
\address{$^{(1)}$Department of Physics, Massachusetts Institute of
Technology, Cambridge MA 02139 \\
$^{(2)}$Department of Physics, Yale University, P.O. Box
208120, New Haven CT 06520-8120 \\
$^{(3)}$Institut f\"ur Theorie der Kondensierten Materie,
Universit\"at Karlsruhe, Postfach 6980, D-76128 Karlsruhe,
Germany}

\date{\today}
\maketitle

\begin{abstract}
In spatial dimensions $d \geq 2$, Kondo lattice models of
conduction and local moment electrons can exhibit a
fractionalized, non-magnetic state (FL$^\ast$) with a Fermi
surface of sharp electron-like quasiparticles, enclosing a volume
quantized by $(\rho_a-1)\mbox{(mod 2)}$, with $\rho_a$ the mean
number of all electrons per unit cell of the ground state. Such
states have fractionalized excitations linked to the deconfined
phase of a gauge theory. Confinement leads to a conventional Fermi
liquid state, with a Fermi volume quantized by $\rho_a \mbox{(mod
2)}$, and an intermediate superconducting state for the $Z_2$
gauge case. The FL$^\ast$ state permits a second order
metamagnetic transition in an applied magnetic field.
\end{abstract}

\maketitle

\begin{multicols}{2}
\narrowtext

The physics of the heavy fermion metals, intermetallic compounds
containing localized spin moments on $d$ or $f$ orbitals and
additional bands of conduction electrons, has been of central
interest in the theory of correlated electron systems for several
decades \cite{hewson,piers1,piers2}. These systems are
conveniently modelled by the much studied Kondo lattice
Hamiltonian, in which there are exchange interactions between the
local moments and the conduction electrons, and possibly
additional exchange couplings between the local moments
themselves. To be specific, one popular Hamiltonian to which our
results apply is:
\begin{eqnarray}
H = &-& \sum_{j,j'} t(j,j') c_{j \sigma}^{\dagger} c_{j' \sigma} +
\frac{1}{2} \sum_{j} J_K (j) \vec{S}_j \cdot c_{j \sigma}
\vec{\tau}_{\sigma\sigma'} c_{j\sigma'} \nonumber \\
&+& \sum_{j<j'} J_{H} (j,j') \vec{S}_j \cdot \vec{S}_{j'}.
\label{h}
\end{eqnarray}
Here the local moments are $S=1/2$ spin $\vec{S}_j$, and the
conduction electrons $c_{j \sigma}$ ($\sigma = \uparrow
\downarrow$) hop on the sites $j$, $j'$ of some regular lattice in
$d$ spatial dimensions with amplitude $t(j,j')$, $J_K
>0$ are the Kondo exchanges ($\vec{\tau}$ are the
Pauli matrices), and explicit short-range Heisenberg exchanges,
$J_H$, between the local moments have been introduced for
theoretical convenience. A chemical potential for the $c_\sigma$
fermions which fixes their mean number at $\rho_c$ per unit cell
of the ground state is implied. We have not included any direct
couplings between the conduction electrons as these are assumed to
be well accounted by innocuous Fermi liquid renormalizations.

For simplicity, we restrict our attention here to non-magnetic
states, in which there is no average static moment on any site
($\langle \vec{S}_j \rangle = 0$), and the spin rotation
invariance of the Hamiltonian is preserved: the $\vec{S}_j$
moments have been `screened', either by the $c_{\sigma}$
conduction electrons, or by their mutual interactions (there is a
natural extension of our results to magnetic states). It is widely
accepted\cite{hewson,piers2,large,oshi1,oshi2,bgg0} that such a
ground state of $H$ is a conventional Fermi liquid (FL) with a
Fermi surface of `heavy' quasiparticles, enclosing a volume,
${\cal V}_{FL}$ determined by the Luttinger theorem:
\begin{equation}
{\cal V}_{FL} = {\cal K}_d [\rho_a\mbox{(mod 2)}]. \label{vfl}
\end{equation}
Here ${\cal K}_d = (2 \pi)^d /(2 v_0)$ is a phase space factor,
$v_0$ is the volume of the unit cell of the ground state, $\rho_a
= n_{\ell} + \rho_c$ is the mean number of all electrons per
volume $v_0$, and $n_{\ell}$ (an integer) is the number of local
moments per volume $v_0$. Note that $\rho_{c,a}$ need not be
integers, and the (mod 2) in (\ref{vfl}) allows neglect of fully
filled bands. In $d=1$, (\ref{vfl}) has been established
rigorously by Yamanaka {\em et al.} \cite{oshi1}. In general $d$,
a non-perturbative argument for (\ref{vfl}), assuming that the
ground state is a Fermi liquid, has been provided by
Oshikawa\cite{oshi2}, who also emphasized that that the Luttinger
theorem can be regarded as a ``quantization'' of ${\cal V}_{FL}$.

The primary purpose of this paper is to show that there exist
non-magnetic, metallic states (FL$^\ast$) in dimensions $d \geq 2$
with a Fermi surface of ordinary $S=1/2$, charge $-e$, sharp
quasiparticles, enclosing a volume
\begin{equation}
{\cal V}_{FL^\ast} = {\cal K}_d [(\rho_a-1)\mbox{(mod 2)}]
\label{vfls}
\end{equation}
over a finite range of parameters. For $n_{\ell}=1$ ${\cal
V}_{FL^\ast}$ is determined by the density of conduction electrons
alone. A number of earlier works\cite{piers3,kagan,bgg} have
considered a Fermi surface of conduction electrons alone,
decoupled in mean-field from the local moments. Here we establish
the conditions under which (\ref{vfls}) characterizes a stable
phase of matter for generic couplings, beyond simple decoupled
models. One of our findings is that any FL$^\ast$ state must be
{\em fractionalized}\cite{sf} {\em i.e.\/} it possesses $S=1/2$
neutral spinon excitations (which are entirely distinct from the
Fermi surface quasiparticles) which carry a charge under a gauge
group which characterises the topological order in the FL$^\ast$
state. We will consider here only the simplest case of a $Z_2$
gauge group\cite{sf,z2}, in which case the $Z_2$ FL$^\ast$ state
possesses a gap to topologically non-trivial `vison' states
\cite{sf,rc,bonesteel} which carry $Z_2$ flux. The connection with
a $Z_2$ (or other) gauge theory explains why the FL$^\ast$ is not
possible in $d=1$: a translationally invariant deconfined phase of
the gauge theory is only present for $d \geq 2$. We will also
discuss the quantum transition between the $Z_2$ FL$^\ast$ state
and the conventional FL state as the exchange couplings are
varied: this transition is preempted by a superconducting state.

We note that the FL$^\ast$ state does not contradict the
non-perturbative computation by Oshikawa \cite{oshi2} of ${\cal
V}_{FL}$; on the contrary, this argument helps establish the
intimate connection between (\ref{vfls}) and topological order.
Oshikawa placed the system on a torus, and considered the
adiabatic evolution of the ground state upon threading a magnetic
flux of $hc/e$ felt by the electrons with spin up (in some basis)
through one of the holes of the torus. For insulating
antiferromagnets with a fractionalized spin liquid ground state (a
resonating valence bond (RVB) state), this procedure connects two
of the topologically distinct states which become degenerate in
the thermodynamic limit in a toroidal geometry
\cite{bonesteel,misguich,rate} {\em i.e.} it connects states with
and without a vison threading the hole of the torus. The FL$^\ast$
state of the Kondo lattice models we are discussing here has a
similar topological order, and the toroidal system has global
vison excitations which are degenerate with the ground state in
the thermodynamic limit. Oshikawa did not consider such
excitations, and only included the electron-like Fermi surface
quasiparticles. Consequently, his argument does not directly apply
to the FL$^\ast$ state, and a modification accounting for vison
excitations shows that the volume ${\cal V}_{FL^\ast}$ is allowed.
In other words, the Fermi volume is still quantized, but
differently from that in a Fermi liquid.

The volume ${\cal V}_{FL}$ is observed in many compounds, and in
particular in those with weak direct exchange $J_H$ between
different local moments. Doniach \cite{doniach} pointed out that
increasing $J_H$ would lead to magnetically ordered states.
However, the effective exchange interactions between the local
moments are strongly frustrated in many common lattices, so that
the magnetic order may be very fragile or entirely absent: it is
these frustrated systems which are favorable candidates for
displaying a non-magnetic FL$^\ast$ state. The generic appearance
of superconductivity in the crossover between the $Z_2$ FL$^\ast$
and FL states is experimentally significant: this may be regarded
as a proposed `mechanism' for superconductivity in heavy fermion
systems, which bears some similarity to the RVB theory \cite{pwa}.
The critical temperature ($T$) for the onset of superconductivity,
$T_c$, can be small.

The $T>0$ behavior of the $Z_2$ FL$^\ast$ state depends on $d$, as
discussed for other fractionalized states in Ref. \cite{sf}. In
$d=3$ there is a finite temperature phase transition associated
with the onset of topological order. This is absent in $d = 2$
where the topological order is present only at $T = 0$. In layered
quasi-two dimensional materials, both types of behavior
(corresponding to two distinct $T = 0$ phases) are possible.

To understand the origin of our results in the context of
(\ref{h}), consider first the limiting case $J_K=0$, when the
$c_\sigma$ fermions and the $\vec{S}_j$ spins are decoupled. While
the $c_\sigma$ fermions will occupy states inside a Fermi surface
enclosing volume ${\cal K}_d [\rho_c \mbox{(mod 2)}]$, there are
two distinct classes of possibilities for a non-magnetic ground
state for the $\vec{S}_j$ spins interacting via $J_H$.

The first is a ground state with confinement of spinons and a unit
cell with $n_{\ell}$ even; this may require breaking of
translational symmetry by the appearance of bond
order\cite{confine}. In this case $\rho_a = \rho_c \mbox{(mod
2)}$, and turning on a finite $J_K$ leads to a FL state, possibly
with co-existing bond order, with the Fermi volume ${\cal V}_{FL}$
equal to that at $J_K = 0$.

The second possibility, of central interest in this paper, is that
$\vec{S}_j$ moments form a fractionalized spin liquid ground state
with $n_{\ell}$ odd \cite{sf,z2,footnl}: this happens on
frustrated lattices, as has been supported by
studies\cite{triangular} on the triangular lattice. A fundamental
property of such a state is its topological stability \cite{sf},
and the associated gap towards creation of vison excitations which
carry unit flux of a $Z_2$ gauge field. The $S=1/2$ spinon
excitations above this state carry a unit $Z_2$ gauge charge. Now
turn on a small $J_K \neq 0$. The key argument of this paper is
that the resulting ground state is smoothly connected to the $J_K
=0$ limit: the quantum numbers of the latter state and its
excitations are topologically protected, the vison gap will
survive for a finite range of $J_K$ values, and perturbation
theory in powers of $J_K$ is non-singular. So we obtain our
advertised FL$^\ast$ state, with a Fermi surface of spin-1/2,
charge $-e$, quasiparticles enclosing the volume ${\cal
V}_{FL^\ast}$ equal to that at $J_K =0$, along with a separate set
of spin-1/2 neutral spinon excitations \cite{demler}. Physically,
each local moment has formed a singlet with another local moment
in an RVB spin liquid state - the Kondo coupling with the
conduction electrons is ineffective in breaking these singlets.
The Fermi surface quasiparticles have a weak residual interaction,
arising from exchanges of pairs of spinons, which could lead to
their pairing in a high angular momentum channel at some very low
$T$: this produces an exotic superconductor which co-exists with a
fractionalized spin liquid \cite{scs} which we will not discuss
further---the superconductivity discussed elsewhere in this paper
is more robust and a qualitatively different state.

As we continue to increase $J_K$, the physics of the Kondo effect
will eventually be manifest: it will become favorable for a local
moment to form a Kondo singlet with the conduction electrons
rather than with other local moments. This may be formalized as
follows, for the case in which the spinons are fermions:
Representing the $\vec{S}_{j}$ moments by $S=1/2$ fermions $f_{j
\sigma}$ ($\vec{S}_{j} = f_{j \sigma}^{\dagger} \vec{\tau}_{\sigma
\sigma'} f_{j \sigma'} /2$ with the single-occupancy constraint
$f_{j \sigma}^{\dagger} f_{j \sigma} = 1$), the formation of Kondo
singlets is signaled by a non-zero hybridization between the
$f_\sigma$ and the $c_\sigma$ fermions. This can be expressed more
precisely in terms of the composite boson fields $B_1 =
f_{\sigma}^{\dagger} c_{\sigma}$ and $B_2 =
\varepsilon^{\sigma\sigma'} f_{\sigma} c_{\sigma'}$, where
$\varepsilon$ is the antisymmetry tensor with
$\varepsilon^{\uparrow\downarrow}=1$. Both of these fields have a
unit $Z_2$ gauge charge, an electromagnetic charge $e$ and are
spin singlet: Condensation of these bosons implies a non-zero
amplitude that a local moment has formed a Kondo singlet with the
conduction electrons. This condensation indicates that the $Z_2$
gauge theory enters a Higgs phase which can also be identified
with a phase in which $Z_2$ charges are confined \cite{fs}.
Moreover, as the spinon pairing amplitude $\langle
\varepsilon^{\sigma\sigma'} f_{\sigma} f_{\sigma'} \rangle$ is
generically non-zero in the small $J_K$ fractionalized phase
\cite{sf}, the condensation of $B_1$ implies condensation of $B_2$
(and vice-versa), and there is only a single $Z_2$ confinement
transition. More importantly, the pairing of the spinons and the
condensation of $B_{1,2}$ implies that the resulting phase also
has pairing of the conduction electrons, and is a superconductor
at $T=0$.

Consider now the behavior when $J_K, t \gg J_H$. In the limit $J_H
=0$, the usual FL state is expected (at least at generic
incommensurate conduction electron density). Turning on a weak
non-zero $J_H$ potentially introduces a weak instability toward
superconductivity, as will be the case in our mean-field theory
below. However, the FL state may still be stabilized by a weak
nearest neighbor repulsive interaction between the conduction
electrons.

The general considerations above can be illustrated by a simple
mean-field computation of the phase diagram of $H$. We applied the
large $N$ method associated with a generalization of $H$ to
Sp($N$) symmetry\cite{tJ} on the triangular lattice. It is
important to note that both the symmetry group and the lattice
have been carefully chosen to allow for a mean field state with
$Z_2$ topological order, stable under gauge fluctuations
\cite{z2}; in particular, there are topologically distinct
mean-field ground states in a toroidal geometry, differing in the
$Z_2$ flux through the holes of the torus. Other
choices\cite{piers3} for the lattice or the symmetry group lead to
mean-field solutions which are generically disrupted by U(1) or
SU(2) gauge fluctuations in $d=2$. We used self-conjugate, fully
antisymmetric (fermionic) representations for the spin states, and
the computations were then similar to earlier work on the $t$-$J$
model \cite{tJ}. For $J_K=0$ and nearest neighbor $J_H$, these
representations yields globally stable solutions in which the
$\vec{S}_j$ spins are paired in fully dimerized states which break
lattice symmetries. As we are not interested in such states here,
we restricted our analysis to saddle points which preserve all
lattice symmetries. Such RVB saddle points can be stabilized by
additional couplings between the local moments; they are also
stable for nearest-neighbor $J_H$ for bosonic spin representions
\cite{sstriangle,tJ}, but these, unfortunately, do not allow a
simple description of the FL state at large $J_K$. It is possible
that the spinons undergo a change from bosonic to fermionic
statistics with increasing $J_K$ within the FL$^\ast$ state, but
this will not be captured by our present mean field theory which
has only fermionic spinons.

\begin{figure}[t]
\epsfxsize=3in \centerline{\epsffile{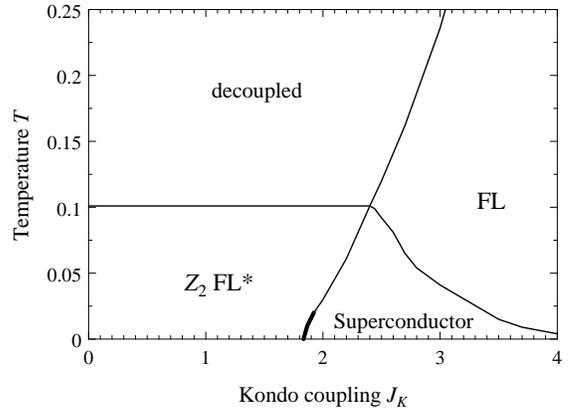}} \caption{Mean
field phase diagram of $H$ on the triangular lattice. We used
fermionic representations of Sp($N$) for the spins, and restricted
attention, by hand, to saddle points which preserve all lattice
symmetries. We had nearest-neighbor $t=1$, $J_H=0.4$, and
$\rho_c=0.7$. The superconducting $T_c$ is exponentially small,
but finite, for large $J_K$, while it is strictly zero for small
$J_K$. Thin (thick) lines are second (first) order transitions.
The transitions surrounding the superconductor will survive beyond
mean field theory, while the others become crossovers.}
\label{fig1}
\end{figure}

The phase diagram  is shown in Fig~\ref{fig1} as a function of
$J_K$ and $T$ for fixed $J_H$, $t$, and $\rho_c$.
In addition to the $Z_2$ FL$^\ast$ and FL states, and an
intermediate superconducting state, whose character we have
already discussed, there is also a high temperature ``decoupled''
state. Here, in the mean field saddle point, the spins are
mutually decoupled from each other, and from the conduction
electrons. This decoupling is, of course, an artifact of the
saddle point, and it points to a regime where all excitations are
incoherent but strongly interacting with each other. For the case
where the superconducting phase is present only at very low
temperatures (as may well be the case beyond mean field theory),
this incoherent regime  represents the quantum-critical region of
the $Z_2$ FL$^\ast$-FL transition. A separate description of this
incoherent quantum critical dynamics was provided by the
large-dimensional saddle point studied by Burdin {\em et al.}
\cite{bgg}, where it was related to the gapless spin liquid state
of Ref.~\onlinecite{sy}.

An interesting $T=0$ quantum phase transition appearing in
Fig~\ref{fig1} is that between the FL$^\ast$ and superconducting
states. As we discussed earlier, this transition is associated
with the condensation of the charge $e$ bosons $B_{1,2}$. A
critical theory of the transition can be written down in terms of
$B_{1,2}$ and the conduction electrons: the methods and resulting
field theory are identical to those discussed in
Ref.~\onlinecite{jinwu}. The renormalization group analysis shows
that the $T=0$ transition can be either first or second order,
depending upon the values of microscopic parameters. The gapped
vison excitations in the FL$^\ast$ state may be detected through
the flux trapping experiments discussed in Ref. \cite{sf2}.
Furthermore, provided the transition is not too strongly first
order, the presence of a critical charge $e$ bosonic mode implies
that the superconducting state in the vicinity of this transition
is a candidate for displaying stable $hc/e$
vortices\cite{ssvortex,sf}.

Interesting physics obtains in the presence of an external uniform
Zeeman magnetic field in the FL$^\ast$ state. As the local moment
and conduction electron systems are essentially decoupled in this
phase, they both respond independently to the magnetic field. If
the spinons are gapped in the fractionalized phase, then there
would be a critical field $B_c$ associated with the onset of
magnetization in the local moment system. Experimentally, this
would be seen as a ``metamagnetic'' transition in the response of
the system to an applied field. Interestingly, this onset
transition could clearly be generically ({\em i.e} without any
fine tuning) second order. Metamagnetic quantum criticality in
strongly correlated systems has been the subject of some recent
experimental\cite{grigera} and theoretical studies\cite{millis},
although accidental fine tuning has been invoked to obtain a
second order transition.

This paper has established that metals with local moments in
dimensions $d \geq 2$ can have non-magnetic ground states
(FL$^\ast$) which are distinct from the familiar heavy Fermi
liquid state (FL). The latter state has a Fermi surface enclosing
a volume ${\cal V}_{FL}$ determined by the density, $\rho_a$, of
both the conduction electrons and local moments; our topologically
ordered FL$^\ast$ state has sharp electron-like excitations on a
Fermi surface enclosing a volume  ${\cal V}_{FL^\ast}$ determined
by $(\rho_a-1)$ (for $n_{\ell}=1$ this is the density of
conduction electrons alone), along with additional
`fractionalized' excitations. In between these FL and FL$^\ast$
states, a plethora of additional states associated with magnetic,
superconducting, and charge order appear possible, along with
non-trivial quantum critical points between them. We believe this
rich phenomenology should find experimental realizations in the
heavy fermion compounds.

We thank A.~Chubukov, P.~Coleman, M.~P.~A.~Fisher, A.~Georges,
P.~A.~Lee, C.~Pepin, and Q.~Si for useful discussions. We were
supported by the MRSEC program of the US NSF under grant number
DMR-9808941 (T.S.), by US NSF Grant DMR 0098226 (S.S.), and by the
DFG through SFB 484 (M.V.). T.S also acknowledges funding from the
NEC Corporation and the hospitality of the Aspen Center for
Physics.

\end{multicols}

\begin{thebibliography}{}

\bibitem{hewson} A.~C.~Hewson, {\em The Kondo Problem to
Heavy Fermions}, (Cambridge University Press, Cambridge, 1993).

\bibitem{piers1} P.~Coleman, C.~P\'epin, Q.~Si, and
R.~Ramazashvili, J. Phys: Condens. Matt. {\bf 13}, 723 (2001).

\bibitem{piers2} P.~Coleman, cond-mat/0206003.

\bibitem{large} H.~Shiba and P.~Fazekas, Prog. Theor. Phys. Supp. {\bf 101}, 403
(1990); K.~Ueda, T.~Nishino, and H. Tsunetsugu, Phys. Rev. B {\bf
50}, 612 (1994); S.~Moukouri and L.~G.~Caron, Phys. Rev. B {\bf
54}, 12212 (1996); P.~Nozi\`{e}res, Eur. Phys. B {\bf 6}, 447
(1998).

\bibitem{oshi1} M.~Yamanaka, M.~Oshikawa, and I.~Affleck, Phys.
Rev. Lett. {\bf 79}, 1110 (1997).

\bibitem{oshi2} M.~Oshikawa, Phys. Rev. Lett. {\bf 84}, 3370 (2000).

\bibitem{bgg0} S.~Burdin, A.~Georges, and
D.~R.~Grempel, Phys. Rev. Lett. {\bf 85},
1048 (2000).

\bibitem{piers3} N.~Andrei and P.~Coleman, Phys. Rev. Lett. {\bf 62}, 595
(1989).

\bibitem{kagan}  Yu. Kagan, K.~A.~Kikoin, and N.~V.~Prokof'ev, Physica B
{\bf 182}, 201-209 (1992).

\bibitem{bgg} S.~Burdin, D.~R.~Grempel, and A.~Georges,
Phys. Rev. B {\bf 66}, 045111 (2002).

\bibitem{sf} T.~Senthil and M.~P.~A.~Fisher, Phys. Rev. B {\bf 62},
7850
(2000); Phys. Rev. B {\bf 63}, 134521 (2001).

\bibitem{z2} N.~Read and S.~Sachdev, Phys. Rev. Lett. {\bf 66},
1773 (1991); X.-G.~Wen, Phys. Rev. B {\bf 44}, 2664 (1991).

\bibitem{rc} S.~Kivelson, Phys. Rev. B {\bf 39}, 259 (1989);
N.~Read and B.~Chakraborty, Phys. Rev. B {\bf 40}, 7133 (1989).

\bibitem{bonesteel} N.~E.~Bonesteel, Phys. Rev. B {\bf 40}, 8954
(1989).

\bibitem{misguich}
G.~Misguich, C.~Lhuillier, M.~Mambrini, and
P.~Sindzingre, Eur. Phys. J. B {\bf 26}, 167 (2002).

\bibitem{rate} For some sample geometries, it is necessary that
the ``adiabatic'' flux insertion rate be faster than the energy
splitting between the nearly degenerate ground states, which is
exponentially small in system size.



\bibitem{doniach} S.~Doniach, Physica B {\bf 91}, 231 (1977).

\bibitem{pwa} P.~W.~Anderson, Science {\bf 235}, 1196 (1987).


\bibitem{confine} N.~Read and S.~Sachdev, Phys. Rev. Lett. {\bf
62},1694 (1989).

\bibitem{footnl} A fractionalized state is also possible for
$n_{\ell}$ even, but in this case ${\cal V}_{FL^\ast} = {\cal
V}_{FL}$, and so the Fermi volume no longer serves as a diagnostic
for fractionalization.

\bibitem{triangular} R.~Moessner and S.~L.~Sondhi, Phys. Rev. Lett.
{\bf 86},
1881 (2001).

\bibitem{demler} Fractionalized states in the model (\protect\ref{h}) were
discussed recently by E.~Demler, C.~Nayak, H.-Y.~Kee, Y.-B.~Kim,
and T.~Senthil, Phys. Rev. B {\bf 65}, 155103 (2002), but they did
not notice that the conduction electrons will retain their bare
quantum numbers at small $J_K$.

\bibitem{scs} This exotic superconductor is the SC$^{*}$ state
introduced in  Ref.~\protect\onlinecite{sf}.

\bibitem{fs} E.~Fradkin and S.~H.~Shenker, Phys. Rev. D {\bf 19}, 3682
(1979).

\bibitem{tJ} S.~Sachdev and N.~Read, Int. J. Mod. Phys. B {\bf 5}, 219
(1991); M.~Vojta, Y.~Zhang, and S.~Sachdev, Phys. Rev. B {\bf 62},
6721 (2000).

\bibitem{sstriangle} S.~Sachdev, Phys. Rev. B {\bf 45}, 12377 (1992).

\bibitem{sy} S.~Sachdev and J.~Ye, Phys. Rev. Lett. {\bf 70}, 3339
(1993).

\bibitem{jinwu} J.~Ye and S.~Sachdev, Phys. Rev. B {\bf 44}, 10173
(1991).


\bibitem{sf2} T.~Senthil and M.~P.~A.~Fisher,
Phys. Rev. Lett. {\bf 86}, 292 (2001).

\bibitem{ssvortex} S. Sachdev, Phys. Rev. B {\bf
45}, 389 (1992); N.~Nagaosa and P.~A.~Lee, Phys. Rev. B {\bf 45},
966 (1992).

\bibitem{grigera} S.~A.~Grigera {\em et al.}, Science {\bf 294},
329 (2001).

\bibitem{millis} A.~J.~Millis {\em et. al.}, Phys. Rev. Lett. {\bf 88}, 217204
(2002).


\end{thebibliography}
\end{document}